\documentstyle[preprint,aps]{revtex}
\tighten
\draft
\widetext
\preprint{YITP-00-32}
\bigskip
\bigskip
\begin{document}
\title{Bulk Supersymmetry and Brane Cosmological Constant}
\medskip
\author{Zurab Kakushadze\footnote{E-mail: 
zurab@insti.physics.sunysb.edu}}
\bigskip
\address{C.N. Yang Institute for Theoretical Physics\\ 
State University of New York, Stony Brook, NY 11794}

\date{June 27, 2000}
\bigskip
\medskip
\maketitle

\begin{abstract} 
{}We consider a recent proposal to solve the cosmological constant problem
within the context of brane world scenarios with infinite volume extra 
dimensions. In such theories bulk can be supersymmetric even if brane
supersymmetry is completely broken. We propose a setup where unbroken bulk 
supersymmetry appears to protect the brane cosmological constant. This
is due to a non-trivial scalar potential in the bulk which implies a 
non-trivial profile for a bulk scalar field. In the presence of the latter
bulk supersymmetry appears to be incompatible with non-vanishing brane
cosmological constant. Moreover, in this setup the corresponding domain wall 
interpolates between an AdS and the Minkowski vacua, so that the weak energy
condition is not violated.
\end{abstract}

{}In the Brane World scenario the Standard Model gauge and matter fields
are assumed to be localized on  
branes (or an intersection thereof), while gravity lives in a larger
dimensional bulk of space-time 
\cite{early,BK,polchi,witt,lyk,shif,TeV,dienes,3gen,anto,ST,BW}. The volume
of dimensions transverse to the branes is 
automatically finite if these dimensions are compact. On the other hand, 
the volume of the transverse dimensions
can be finite even if the latter are non-compact. In particular, this can be
achieved by using \cite{Gog} warped compactifications \cite{Visser} which
localize gravity on the brane. A concrete
realization of this idea was given in \cite{RS}.

{}Recently it was pointed out in
\cite{DGP0,witten} that, in theories where extra dimensions
transverse to a brane have infinite volume\cite{GRS,CEH,DVALI,DGP1,DGP}, 
the cosmological constant on the
brane might be under control even if brane supersymmetry is completely
broken. The key point here is that even if supersymmetry breaking on the
brane does take place, it will not be transmitted to the bulk as the volume 
of the extra dimensions is infinite \cite{DGP0,witten}. Thus, at least in 
principle, one should be able to control some of the properties of the bulk
with the unbroken bulk supersymmetry. In particular, vanishing of the
bulk cosmological constant need not be unnatural\footnote{Note that {\em a
priori} we could have negative cosmological constant consistent with
bulk supersymmetry. However, in the presence of supersymmetry various ways 
are known for ensuring vanishing of the bulk cosmological constant without any
fine-tuning.}.

{}Then the ``zeroth-order'' argument goes as follows \cite{DGP0,witten}.
Let us for definiteness focus on the case of the codimension one brane
embedded in $D$-dimensional space-time. At
least naively, at large (enough) distances, which are precisely relevant 
for the discussion of the cosmological constant, the theory is expected 
to become 
$D$-dimensional. In particular, the laws of gravity, such as Newton's law, 
are expected to become $D$-dimensional at such distances. If so, a brane 
world observer would then really be
measuring the $D$-dimensional (and not $(D-1)$-dimensional) cosmological
constant, which vanishes by bulk supersymmetry. One therefore might 
expect that
the cosmological constant on the brane might somehow also vanish
regardless of brane supersymmetry.

{}However, as was recently pointed out in \cite{zura}, this argument need not
always apply. In particular, it was pointed out in \cite{zura} that in a 
concrete model of \cite{DGP}, which we will refer to as the 
Dvali-Gabadadze-Porrati model, bulk supersymmetry is perfectly compatible with
positive brane cosmological constant, that is, the former does not control the
latter in this model. This, in turn, appears to be related to the fact that,
in the presence of positive cosmological constant on the brane, the bulk 
graviton spectrum has a mass gap in this model, so that below the corresponding
energy scale, which is set precisely by the brane cosmological constant, the
theory becomes effectively $(D-1)$-dimensional.

{}Here we would like to mention another potential difficulty with the
infinite volume brane world scenarios. Thus, as was pointed out in 
\cite{witten,CEH}, if the bulk asymptotically approaches the Minkowski space
on both sides of a (codimension one) brane, then the weak energy condition
is necessarily violated if the brane cosmological constant vanishes. In 
particular, such a scenario would involve a negative tension brane which
suffers from the presence of world-volume ghosts (as long as the brane 
cosmological constant vanishes). This problem can be avoided if the bulk is
completely flat, that is, the bulk cosmological constant vanishes identically.
We then essentially arrive at the Dvali-Gabadadze-Porrati model, but, as we 
have already mentioned, in the latter bulk supersymmetry does not protect the
brane cosmological constant.
 
{}In this paper we would like to propose a setup which appears to circumvent
both of the aforementioned difficulties. In particular, unbroken bulk 
supersymmetry does seem to protect the brane cosmological constant in this
setup. Moreover, weak energy condition is preserved as on one side of the
brane the bulk approaches an AdS space, while on the other side it approaches 
the Minkowski space. Note that in this case the volume of the transverse
dimension is still infinite, and bulk supersymmetry is still unbroken even
if brane supersymmetry is. (A more detailed discussion of this setup,
including properties of brane world gravity in this context, will be
presented elsewhere \cite{new}.) 

{}In fact, in this paper we will consider a concrete
model with the aforementioned properties. The (relevant part of the) 
action for this model is given by:
\begin{equation}\label{action}
 S={\widehat M}_P^{D-3}\int_\Sigma d^{D-1} x \sqrt{-{\widehat G}}\left[
 {\widehat R}-{\widehat\Lambda}\right] +
 M_P^{D-2}
 \int d^D x \sqrt{-G} \left[R-{4\over{D-2}}(\nabla\phi)^2-V(\phi) \right]~.
\end{equation}
For calculational convenience we will keep the number of space-time
dimensions $D$ unspecified.
In (\ref{action}) ${\widehat M}_P$ is (up to a normalization factor - see 
below) the $(D-1)$-dimensional (reduced) Planck scale, while $M_P$ is the 
$D$-dimensional one. The $(D-1)$-dimensional hypersurface $\Sigma$, which we
will refer to as the brane, is the $y=y_0$ slice of the $D$-dimensional 
space-time,
where $y\equiv x^D$, and $y_0$ is a constant. Next, 
\begin{equation}
 {\widehat G}_{\mu\nu}\equiv{\delta_\mu}^M {\delta_\nu}^N G_{MN}
 \Big|_{y=y_0}~,
\end{equation} 
where the capital Latin indices $M,N,\dots=1,\dots,D$, while the Greek
indices $\mu,\nu,\dots=1,\dots,(D-1)$. The quantity ${\widehat\Lambda}$ is
the brane tension. More precisely, there might be various (massless and/or 
massive) fields (such
as scalars, fermions, gauge vector bosons, {\em etc.}), which we
will collectively denote via $\Phi^i$, localized on the brane. Then ${\widehat
\Lambda}={\widehat\Lambda}(\Phi^i,\nabla_\mu\Phi^i,\dots)$ generally depends
on the vacuum expectation values of these fields as well as their derivatives.
In the following we will assume that the expectation values of the $\Phi^i$
fields are dynamically determined, independent of the coordinates 
$x^\mu$, and consistent with $(D-1)$-dimensional general covariance. 
The quantity ${\widehat\Lambda}$ is then a constant which we identify 
as the brane tension. 
The bulk fields are given by the metric $G_{MN}$, a single
real scalar field $\phi$, as well as other fields (whose expectation values
we assume to be vanishing) which would appear in a concrete supergravity 
model (for the standard values of $D$). 
Finally, let us note that in the action (\ref{action}) we assume 
${\widehat \Lambda}$ to be independent of $\phi$, that is, the bulk scalar
$\phi$ does not couple to the brane.

{}The model defined in (\ref{action}) is a generalization of the 
Dvali-Gabadadze-Porrati model recently proposed in \cite{DGP}. In fact, the
difference between the two models (on top of the straightforward 
generalization that we do not 
{\em a priori} assume that the brane is tensionless) 
is the presence of the bulk scalar field which we will assume to have a 
non-trivial scalar potential $V(\phi)$ (further assumptions on $V(\phi)$
will be discussed below). In fact, the latter will play the key role in
what follows. 

{}Before we turn to our main point, let us briefly comment on the 
$\sqrt{-{\widehat G}} {\widehat R}$ term in the brane world-volume action.
Typically such a term is not included in discussions of various brane world
scenarios (albeit usually the $-\sqrt{-{\widehat G}} {\widehat \Lambda}$ 
term is). 
However, as was pointed out in \cite{DGP}, even if such a term is absent at the
tree level, as long as the 
brane world-volume theory is not conformal, it will typically be
generated by quantum loops of other fields localized on the 
brane (albeit 
not necessarily with the desired sign, which, nonetheless, appears to be as
generic as the opposite one). This is an important observation, which allows 
to reproduce the $(D-1)$-dimensional Newton's law on the brane in the setup 
we discuss in this paper.

{}To proceed further, we will need equations of motion following from the
action (\ref{action}). Here we are interested in studying possible solutions
to these equations which are consistent with $(D-1)$-dimensional general 
covariance. That is, we will be looking for solutions with the warped
metric of the following form:
\begin{equation}\label{warped}
 ds_D^2=\exp(2A)ds_{D-1}^2+dy^2~,
\end{equation}
where the warp factor $A$, which is a function of $y$,
is independent of the coordinates
$x^\mu$, and the $(D-1)$-dimensional interval is given
by
\begin{equation}\label{D-1}
 ds_{D-1}^2={\widetilde g}_{\mu\nu}dx^\mu dx^\nu~,
\end{equation}  
with the $(D-1)$-dimensional metric 
${\widetilde g}_{\mu\nu}$ independent of $y$. With this 
ans{\"a}tz, we have the following
equations of motion for $\phi$ and $A$:
\begin{eqnarray}\label{phi''d}
 &&{8\over {D-2}}\left[\phi^{\prime\prime}+(D-1)A^\prime\phi^\prime\right]-
 V_\phi=0~,\\
 \label{phi'A'd}
 &&(D-1)(D-2)(A^\prime)^2-{4\over{D-2}}(\phi^\prime)^2+V-
 {{D-1}\over{D-3}}{\widetilde \Lambda}\exp(-2A)=0~,\\
 \label{A''d}
 &&(D-2)A^{\prime\prime}+{4\over {D-2}}(\phi^\prime)^2+{1\over {D-3}}
 {\widetilde \Lambda}\exp(-2A)+{1\over 2} L f\delta(y-y_0)=0~.
\end{eqnarray}
Here 
\begin{equation}
 f\equiv{\widehat \Lambda}- {\widetilde \Lambda}\exp[-2A(y_0)]~.
\end{equation}
The scale $L$, defined as
\begin{equation}
 L\equiv {\widehat M}_P^{D-3}/M_P^{D-2}~,
\end{equation}
plays the role of the crossover distance scale below which gravity is
effectively $(D-1)$-dimensional, while above this scale it becomes 
$D$-dimensional\footnote{Here we should note that, as was pointed out in 
\cite{zura}, this need not always be the case.}. Next, ${\widetilde \Lambda}$ 
is independent of $x^\mu$ and $y$. In fact, it 
is nothing but the cosmological constant of the $(D-1)$-dimensional manifold,
which is therefore an Einstein manifold, corresponding to the hypersurface
$\Sigma$. Our normalization of ${\widetilde\Lambda}$ is such that
the $(D-1)$-dimensional metric ${\widetilde g}_{\mu\nu}$ satisfies
Einstein's equations:
\begin{equation}
 {\widetilde R}_{\mu\nu}-{1\over 2}{\widetilde g}_{\mu\nu}
 {\widetilde R}=-{1\over 2}
{\widetilde g}_{\mu\nu}{\widetilde\Lambda}~.
\end{equation}

{}Here we note that in the bulk (that is, for $y\not=y_0$) one of the
second order equations is automatically satisfied once the first
order equation (\ref{phi'A'd}) as well as the other second order equation are
satisfied. As usual, this is a consequence of Bianchi identities.

{}Note that by rescaling the coordinates $x^\mu$ on the brane we can always
set $\exp[A(y_0)]=1$. Then the $(D-1)$-dimensional Planck scale is simply
${\widehat M}_P$. Note that the above system of equations has smooth solutions
for $f=0$, that is, if the brane cosmological constant and brane
tension are equal: 
\begin{equation}
 {\widetilde \Lambda}={\widehat\Lambda}~.
\end{equation}
In particular, in these solutions $\phi$ and $A$ as well as their derivatives 
$\phi^\prime$ and $A^\prime$ are smooth. 

{}Before we discuss these solutions, let us make the following remarks. 
For $f\not=0$
there generically (subject to the appropriately chosen scalar potential) 
will exist additional solutions with continuous $\phi^\prime$ but
discontinuous $A^\prime$. However, for $f>0$ such solutions necessarily 
have a tachyonic $\phi$ mode localized on the brane \cite{new}. On the other 
hand, for $f<0$ there are no tachyonic $\phi$ modes localized on the brane, in
fact, the corresponding modes are massive \cite{new}. 
However, for, say, ${\widetilde
\Lambda}=0$ we then have a negative tension brane as ${\widehat\Lambda}<0$,
which suffers from world-volume ghosts. In fact, to possibly avoid world-volume
ghosts, we must assume that the brane cosmological constant is positive, and
then there is a non-vanishing lower bound on ${\widetilde\Lambda}$ \cite{new}.
Finally, 
let us comment on our assumption that ${\widehat\Lambda}$ is independent of 
$\phi$. {\em A priori} one can relax this assumption, which would only modify
(\ref{phi''d}) as follows:
\begin{equation}
 {8\over {D-2}}\left[\phi^{\prime\prime}+(D-1)A^\prime\phi^\prime\right]-
 V_\phi-f_\phi\delta(y-y_0)=0~.
\end{equation}
Now, let $\phi_0\equiv \phi(y_0)$. If $f_\phi(\phi_0)\not=0$, $\phi^\prime$
is then necessarily discontinuous at $y=y_0$. In this case the bulk 
$\phi$ modes responsible for decoupling the ghost-like trace $h^\mu_\mu$ modes
in the bulk \cite{COSM}
cannot be defined, so the latter persist making the bulk theory 
inconsistent \cite{new}. We, therefore, must assume that $f_\phi(\phi_0)=0$,
albeit $f(\phi)$ could still {\em a priori} 
have non-trivial $\phi$ dependence. Even so, the
mass squared of the $\phi$ mode localized on the brane is still given (up to a
positive multiplicative constant) by $-f(\phi_0)$ (and not by $f_{\phi\phi}
(\phi_0)$) \cite{new}, which therefore must be non-negative. These as well as
other subtleties will be discussed in more detail in \cite{new}. However, they
will not be too important for our discussion in this paper. We would, however,
like to 
end this detour with the following remark. If ${\widehat\Lambda}$ does depend
non-trivially on $\phi$, even if $f(\phi_0)=f_\phi(\phi_0)=0$, then generically
we expect that the kinetic term for $\phi$ localized on the brane will be 
generated just as it happens for the brane graviton mode. This would then
imply that we have a massless scalar field localized on the brane, which is
generically (unless for some reason it does not have non-derivative couplings 
to the matter fields localized on the brane)
expected to be in conflict with tight experimental bounds on the
existence of the ``fifth force''. It is precisely to avoid potential 
complications with this issue why we assumed in the very beginning that the
$\phi$ field does not couple to the brane. Here we would like to emphasize 
that, if the $\phi$ field does not couple to the brane, then there is
no $\phi$ zero-mode (which should not be confused with a possible 
translational zero mode localized on the brane)
localized on the brane - the zero mode corresponding to
the broken (by the smooth domain wall) 
translational invariance is eaten by the graviphoton (arising
in the reduction of the $D$-dimensional graviton in terms of 
$(D-1)$-dimensional fields) in the corresponding Higgs mechanism discussed
in \cite{COSM}.

{}Let us now discuss possible solutions of the above system of equations 
(\ref{phi''d}), (\ref{phi'A'd}) and ({\ref{A''d}) for
$f=0$. To obtain an infinite volume solution, let us assume that the scalar
potential has one AdS minimum located at $\phi=\phi_-$ and one Minkowski
minimum located at $\phi=\phi_+$ (without loss of generality we will assume
that $\phi_+>\phi_-$). Moreover, let us assume that there are no other 
extrema except for a dS maximum located at $\phi=\phi_*$, where $\phi_-<\phi_*
<\phi_+$, such that $V(\phi_*)\gg V(\phi_+)-V(\phi_-)=|V(\phi_-)|$. This latter
condition is necessary to sufficiently suppress the probability for nucleation
of AdS bubbles in the Minkowski vacuum, which could otherwise destabilize
the background \cite{okun}. Then (subject to certain conditions on the behavior
of $V(\phi)$ near the minima $\phi_\pm$ - see \cite{zura,new} for details)
we can have smooth domain walls interpolating between the two vacua. In fact,
for ${\widetilde\Lambda}=0$ we have $\phi(y)\rightarrow\phi_\pm$ as 
$y\rightarrow\pm\infty$. On the other hand, for ${\widetilde\Lambda}>0$ we have
$\phi(y)\rightarrow\phi_+$ as $y\rightarrow+\infty$, while 
$\phi(y)\rightarrow\phi_-$ as $y\rightarrow y_-$, where $y_-<y_0$ is finite 
(here for definiteness we have assumed that the domain wall approaches the 
Minkowski vacuum as $y\rightarrow+\infty$). A more detailed discussion of
these domain walls will be given in \cite{new}. However, what is important
for us here is that the warp factor $A$ goes to $-\infty$ as $\phi\rightarrow
\phi_-$ (if ${\widetilde
\Lambda}=0$, then $A$ goes to $-\infty$ linearly with $|y|$, while if 
${\widetilde \Lambda}>0$, then $A\sim\ln(y-y_-)$ as $y\rightarrow y_-$). On 
the other hand, if ${\widetilde \Lambda}=0$, then $A$ goes to a constant
as $y\rightarrow+\infty$, while if ${\widetilde\Lambda}>0$, then $A$ grows
logarithmically with $y$. In both cases the volume of the extra dimension is
infinite as the integral 
\begin{equation}
 \int dy \exp[(D-1)A]
\end{equation}
diverges. Moreover, there are no quadratically normalizable bulk graviton 
modes. Rather, for ${\widetilde\Lambda}=0$ we have a continuum of plane-wave
normalizable bulk modes (with mass squared $m^2\geq 0$), while for
${\widetilde\Lambda}>0$ we have a mass gap in the bulk graviton spectrum,
and the plane-wave normalizable modes are those with $m^2>m_*^2$, where
$m_*^2\sim{\widetilde\Lambda}$ \cite{zura}. Thus, without any additional 
assumptions consistent solutions with vanishing as well as positive 
brane cosmological constant exist for such potentials.
   
{}We would now, however, 
like to point that, as long as the scalar potential $V(\phi)$ is
non-trivial, bulk supersymmetry is incompatible with non-zero brane
cosmological constant. Indeed, this immediately follows from the bulk Killing
spinor equations (which follow from the requirement that variations of the
superpartner $\lambda$ of $\phi$ and the gravitino $\psi_M$ vanish
under the corresponding supersymmetry transformations):
\begin{eqnarray}\label{Killing1}
 &&\left[\Gamma^M\partial_M\phi-\alpha W_\phi\right]\varepsilon=0~,\\
 &&\left[{\cal D}_M-{1\over 2}\beta W \Gamma_M\right] \varepsilon=0~.
 \label{Killing2}
\end{eqnarray}
Here $\varepsilon$ is a Killing spinor, $\Gamma_M$ are $D$-dimensional
Dirac gamma matrices satisfying
\begin{equation}
 \left\{\Gamma_M~,~\Gamma_N\right\}=2 G_{MN}~,
\end{equation}
and ${\cal D}_M$ is the covariant derivative
\begin{equation}
 {\cal D}_M\equiv \partial_M +{1\over 4}\Gamma_{AB}{\omega^{AB}}_M~.
\end{equation}
The spin connection ${\omega^{AB}}_M$ is defined via the vielbeins
${e^A}_M$ in the usual way (here the capital Latin indices $A,B,\dots=1,\dots,
D$ are lowered and raised with the $D$-dimensional Minkowski metric
$\eta_{AB}$ and its inverse, while
the capital Latin indices $M,N,\dots=1,\dots,D$ are lowered and raised with
the $D$-dimensional metric $G_{MN}$ and its inverse). Furthermore,
\begin{equation}
 \Gamma_{AB}\equiv{1\over 2}\left[\Gamma_A~,~\Gamma_B\right]~,
\end{equation} 
where $\Gamma_A$ are the constant Dirac gamma matrices satisfying
\begin{equation}
 \left\{\Gamma_A~,~\Gamma_B\right\}=2 \eta_{AB}~.
\end{equation}
Finally, $W$ is
interpreted as the superpotential in this context.

{}Next, we would like to study the above Killing spinor equations in the
warped backgrounds of the form (\ref{warped}):
\begin{eqnarray}\label{Kill1}
 &&\left[\Gamma_D\phi^\prime-\alpha W_\phi\right]\varepsilon=0~,\\
 \label{Kill2}
 &&\varepsilon^\prime-{1\over 2}\beta W\Gamma_D\varepsilon=0~,\\
 \label{Kill3}
 &&{\widetilde {\cal D}}_\mu \varepsilon+{1\over 2}\exp(A)
 {\widetilde \Gamma}_\mu\left[A^\prime\Gamma_D-
 \beta W\right]\varepsilon=0~.
\end{eqnarray}
Here ${\widetilde{\cal D}}_\mu$ is the $(D-1)$-dimensional covariant derivative
corresponding to the metric ${\widetilde g}_{\mu\nu}$,
${\widetilde \Gamma}_\mu$ are the $(D-1)$-dimensional Dirac gamma
matrices satisfying
\begin{equation}
 \left\{{\widetilde \Gamma}_\mu~,~{\widetilde \Gamma}_\nu\right\}=
 2{\widetilde g}_{\mu\nu}~.
\end{equation}
Also, note that $\Gamma_D$, which is the $D$-dimensional Dirac
gamma matrix $\Gamma_M$ with $M=D$ (that is, the Dirac gamma matrix
corresponding to the $x^D=y$ direction), is constant in this background.
Finally, 
\begin{equation}
 \alpha\equiv\eta{\sqrt{D-2}\over 2}~,~~~\beta\equiv-\eta
 {2\over (D-2)^{3/2}}~,
\end{equation}
where $\eta=\pm 1$.

{}From (\ref{Kill1}), (\ref{Kill2}) and (\ref{Kill3}) it is clear that we
can only have Killing spinors of one helicity (w.r.t. $\Gamma_D$), which 
without loss of generality can be chosen to be positive. Then $\phi$ and $A$
must satisfy the first order (BPS) equations
\begin{eqnarray}
 &&\phi^\prime=\alpha W_\phi~,\\
 &&A^\prime=\beta W~,
\end{eqnarray}
which are compatible with the system of equations (\ref{phi''d}), 
(\ref{phi'A'd}) and (\ref{A''d}) if and only if ${\widetilde\Lambda}=0$,
and the scalar potential is given by
\begin{equation}
 V=W_\phi^2-\gamma^2 W^2~,
\end{equation}
where 
\begin{equation}
 \gamma\equiv {2\sqrt{D-1}\over{D-2}}~.
\end{equation}
Thus, bulk supersymmetry (note that the domain wall solution preserves
$1/2$ of the supersymmetries corresponding to the minima of $V$) is preserved
if and only if the brane cosmological constant vanishes. Note that the key 
difference between this setup and the Dvali-Gabadadze-Porrati model (where
all bulk supersymmetries are preserved by solutions with any non-negative
${\widetilde
\Lambda}$ \cite{zura}) is that in the present context $\phi^\prime$ is not 
identically zero. In fact, if we take the identically vanishing potential 
$V\equiv 0$, then we will have solutions with $\phi={\rm const.}$, which have 
the same properties as those in the Dvali-Gabadadze-Porrati model.

{}We therefore conclude that, 
even if brane supersymmetry is broken in the above setup,
bulk supersymmetry, which remains unbroken as the volume of the 
transverse dimension is infinite, ensures that the brane
cosmological constant still vanishes in the model defined in (\ref{action}).
This, in particular, implies that even with broken brane supersymmetry the
brane tension must for consistency remain zero. There seems to be more than
one way for achieving this, but we will mention only one. Assume that there
is a $(D-2)$-form antisymmetric gauge field localized on the brane. Its field
strength, which is a $(D-1)$-form, can then acquire a 
non-zero expectation value
without breaking Poincar{\'e} invariance on the brane. In fact with the 
appropriate sign for its kinetic term, it can cancel other contributions to the
brane tension such as those due to quantum loops involving the fields 
localized on the brane. (Note that this does no longer appear to be the
usual fine-tuning of the cosmological constant in, say, four dimensions as
the brane cosmological constant as well as the brane tension are now forced
to vanish due to bulk supersymmetry.) Next,
let us assume that the latter contributions to the
brane tension add up to a positive number. Then the kinetic term for the 
three-form potential would have to have a ghost-like sign instead of the usual
one. This, however, is not a problem as the three-form has no propagating
degrees of freedom. Nonetheless, adding such a three-form might appear a bit
{\em ad hoc}. However, here we can adapt the idea of \cite{gabad} (there it was
intended to generate negative contributions to the five-dimensional bulk
cosmological constant using the four-form gauge field with a ghost-like 
kinetic term) that such a three-form field is dual to an auxiliary scalar
component of an ${\cal N}=1$ chiral superfield in four dimensions (here we
are assuming a brane world scenario with a 3-brane embedded in a 5-dimensional
bulk). Such a dualization was explicitly shown to be possible in the context
of ${\cal N}=1$ supergravity in flat $3+1$ dimensions in \cite{ovrut}, and 
leads to the three-form supergravity where the kinetic term for the
three-form has precisely the aforementioned ghost-like sign. Thus,
in this context, perhaps somewhat ironically, quantum corrections to the brane
tension would be canceled by the dual of (a component of) an auxiliary 
$F$-field.

{}We would like to end our discussion here with some remarks. In particular,
we would like to address the question of how the mechanism for vanishing
of the brane cosmological constant described above could possibly fail. One
possibility is that the above discussion does not at all imply that the
brane cosmological constant vanishes even if brane supersymmetry is broken but
rather that brane supersymmetry can never be broken in such a 
scenario\footnote{This possibility was originally 
pointed out to me by Gia Dvali and 
Gregory Gabadadze in a somewhat different context.}. It is difficult to argue
for or against this possibility at least for the following reason. Suppose
on the brane we have a four-dimensional 
low energy effective field theory which is classically
${\cal N}=1$ supersymmetric but quantum mechanically (non-perturbatively)
breaks supersymmetry. In this case the above mechanism would seem to imply
that the brane cosmological constant would have to vanish. However, one could
ask whether such a brane world model can be consistently constructed (with all
possible anomalies canceled and all that) within the context of, say, 
supergravity or string theory (including an explicit self-consistent 
realization of the brane itself)\footnote{Here we would like to emphasize
that bulk supersymmetry does not {\em a priori} imply that brane must be
supersymmetric as the volume of the extra dimension is infinite 
\cite{DGP0,witten}.}. 
At present it is unclear what the answer
to this question should be. 

{}Another possibility is related to one of the
observations of \cite{COSM} that if we have warped backgrounds with 
$A\rightarrow -\infty$, then one must be cautious about higher derivative
terms in the bulk action. Had we considered singular compactifications where
the bulk curvature diverges on the side where we assumed AdS minimum, then 
higher derivative terms might indeed have made unclear what is happening with
the warp factor \cite{COSM}. 
However, in this case we have constant curvature at 
$y\rightarrow-\infty$, so as long as $|V(\phi_-)|\ll
M_*^2$, where $M_*$ is the cut-off scale for higher derivative terms in the
bulk action, then contributions of such terms as far as the domain wall 
solution is concerned appear to be under control \cite{COSM}. 
However, as was also pointed out in \cite{COSM}, higher curvature terms in such
warped compactifications lead to delocalization of gravity. In the above 
context delocalization of gravity does not {\em a priori} appear to be 
a pressing issue as bulk gravity is not localized to begin with. However,
inclusion of higher derivative terms of, say, the form
\begin{equation}
 \zeta\int d^Dx \sqrt{-G} R^k
\end{equation}
into the bulk action would produce terms of the form \cite{COSM}
\begin{equation}
 \zeta \int d^{D-1}x dy\exp[(D-1-2k)A]\sqrt{-{\widetilde g}}{\widetilde R}^k~.
\end{equation}
Assuming that $A$ goes to $-\infty$ at $y\rightarrow-\infty$, for large enough
$k$ the factor $\exp[(D-1-2k)A]$ 
diverges too fast so that at the end of the day
there might no longer be any plane-wave normalizable bulk modes. Such a 
background, at least naively, could be suspected to be either 
somewhat inconsistent, or inadequate for solving the cosmological constant 
problem with bulk supersymmetry. Indeed, for such backgrounds it is not clear
what bulk supersymmetry means and how it can possibly protect brane 
cosmological constant - after all, the entire idea of using bulk supersymmetry 
in this context is based on the observation that the theory becomes 
$D$-dimensional in the infra-red, and if there are no normalizable bulk modes,
then the extra dimension effectively seems to actually disappear.

{}A possible way around this difficulty might be that all the higher curvature
terms should come in ``topological'' combinations (corresponding to Euler
invariants such as the Gauss-Bonnet term \cite{Zwiebach,Zumino}) 
so that their presence does not
modify the $(D-1)$-dimensional propagator for the bulk graviton modes. That is,
even though such terms are multiplied by diverging powers of the warp factor,
they are still harmless. One could attempt to justify the fact that higher
curvature bulk terms must arise only in such combinations by the fact that
otherwise the bulk theory would be inconsistent to begin with due to the
presence of ghosts. However, it is not completely obvious whether it is
necessary to have only such combinations to preserve unitarity. Thus, in
a non-local theory such as string theory unitarity might be preserved,
even though at each higher derivative order there are non-unitary terms, due
to a non-trivial cancellation between an infinite tower of such terms.

{}The third possibility is more prosaic. {\em A priori} 
there is no guarantee that potentials
of the aforementioned type, where one has one AdS and one Minkowski vacua,
are compatible with supersymmetry in a given dimension $D$. Thus, various
attempts to construct 
smooth supersymmetric domain walls 
interpolating between two AdS vacua
in the context of, say, 
$D=5$ ${\cal N}=2$ gauged supergravity have not been
successful so far (see, {\em e.g.}, 
\cite{BC,KL,BG}\footnote{Note, however, that such domain walls 
have been constructed within the framework of $D=4$ ${\cal N}=1$ 
supergravity - see \cite{cvet} for a review.}). In fact, as
was argued in
\cite{ZW,GL}, on general grounds potentials with more then one AdS minima 
are not expected to exist in $D=5$ ${\cal N}=2$ gauged 
supergravity\footnote{For a recent analysis of general $D=5$ ${\cal N}=2$
gauged supergravity models, see \cite{ceresole}.}. 
One argument leading to such a conclusion is based on the 
observation that to have two adjacent AdS vacua, the superpotential $W$ must
change sign between the corresponding values of the scalar $\phi$. According to
\cite{GL} this, however, is inconsistent with supersymmetry. Note that in the
case of one AdS and one Minkowski vacua the superpotential does not change 
sign. An example of such a superpotential is
\begin{equation}
 W=\xi\left[v^2\phi-{1\over 3}\phi^3-{2\over 3} v^3\right]~,
\end{equation}
where we will assume that $|v|\ll 1$. Then the condition $|V(\phi_-)|\ll
V(\phi_*)$ is satisfied, where $\phi_\pm=\pm v$. The domain wall solution in
this case is given by
\begin{eqnarray}
 &&\phi(y)=v\tanh(\alpha\xi v(y-y_1))~,\\
 &&A(y)={2\beta\over3\alpha}v^2 \left[\ln(\cosh(\alpha\xi v(y-y_1)))-
 {1\over 4}{1\over{\cosh^2(\alpha\xi v(y-y_1))}}\right]-
 {2\beta\over 3}\xi v^3(y-y_1)+C~,
\end{eqnarray}  
where $y_1$ and $C$ are integration constants. Thus, at least the argument of
\cite{GL} does not obviously rule out such potentials.

{}Here we would like to point out that even the requirement of having local
minima can in principle be relaxed (which might be useful as far as embedding
such a scenario in supergravity). Thus, one can consider potentials with, say,
no minima at all, but such that they asymptotically approach AdS and Minkowski 
vacua in a runaway fashion. Thus, consider the superpotential
\begin{equation}
 W=\xi\left[1-\tanh(a\phi)\right]~.
\end{equation}
Note that the corresponding scalar potential has no local minima, but has one
maximum. Moreover, at $\phi\rightarrow-\infty$ $V$ approaches an AdS vacuum
with $V(-\infty)=-4\gamma^2\xi^2$, whereas at $\phi\rightarrow+\infty$
$V$ approaches the Minkowski vacuum (for definiteness we are assuming $a>0$). 
The condition $|V(\phi_-)|\ll
V(\phi_*)$ is satisfied provided that $a\gg\gamma$.
The domain wall solution in
this case is given by
\begin{eqnarray}
 &&\sinh(2a\phi)+2a\phi=-4\xi\alpha a^2(y-y_1)~,\\
 &&A=C+{\beta\over4\alpha a^2}\left[\exp(-2a\phi)-2a\phi\right]~.
\end{eqnarray}  
It is not clear whether such potentials can arise in, say, $D=5$ 
${\cal N}=2$ supergravity, but this question is beyond the scope of this 
paper. 

{}An interesting feature of runaway type of potentials is that $A$ no longer
goes to a constant on the Minkowski side of in the corresponding domain wall 
solutions but increases logarithmically. One of the implications of this fact
is that the bulk graviton zero mode is no longer plane-wave normalizable,
albeit the massive bulk modes are \cite{new}. 
The absence of a normalizable bulk zero
mode might affect the way gravity is modified at large distances. In
particular, it is not completely obvious whether the crossover scale between
the $(D-1)$- and $D$-dimensional laws of gravity is still given by $L$. 
It would be interesting to understand whether this could relax the 
phenomenological
upper bound on the five-dimensional Planck scale $M_P$
which in the presence of a 
normalizable bulk zero mode was argued in \cite{DGP} to be
$\sim{\rm GeV}$ or so. 
These and other issues will be discussed in more detail
in \cite{new}.

{}I would like to thank Gregory Gabadadze for an extremely 
valuable discussion.
This work was supported in part by the National Science Foundation.
I would also like to thank Albert and Ribena Yu for financial support.

\end{document}